\documentclass[12pt]{article}
\usepackage{authblk}
\usepackage{graphicx}
\usepackage{siunitx}
\usepackage{amsmath}
\usepackage{amssymb}
\usepackage[a4paper, margin=1in]{geometry}
\usepackage{setspace}
\usepackage{parskip}
\usepackage{fancyvrb}
\usepackage{hyperref}
\usepackage{url}
\usepackage{natbib}

\setcitestyle{numbers}

\newcommand{\Izz}{I_{zz}}
\newcommand{\dist}{r}
\newcommand{\fddot}{\ddot{f}}
\newcommand{\fdot}{\dot{f}}
\newcommand{\pmp}{m_p}

\begin{document}

\title{%
What are neutron stars made of? \\ Gravitational waves may reveal the answer%
\footnote{This essay was awarded 5th Place in the Gravity Research Foundation 2023 Awards for Essays on Gravitation.}%
}
\author[1,2]{Neil Lu\thanks{\texttt{Neil.Lu@anu.edu.au}}}
\author[1,2]{Susan M. Scott\thanks{\texttt{Susan.Scott@anu.edu.au}}}
\author[1,2]{Karl Wette\thanks{\texttt{Karl.Wette@anu.edu.au}  \hspace{1 mm} corresponding author}}
\affil[1]{Centre for Gravitational Astrophysics, The Australian National University, Canberra ACT 2601, Australia}
\affil[2]{Australian Research Council Centre of Excellence for Gravitational Wave Discovery (OzGrav)}
\date{\today}

\maketitle

\doublespacing

\begin{abstract}
Neutron stars are one of the most mysterious wonders in the Universe. Their extreme densities hint at new and exotic physics at work within. Gravitational waves could be the key to unlocking their secrets. In particular, a first detection of gravitational waves from rapidly-spinning, deformed neutron stars could yield new insights into the physics of matter at extreme densities and under strong gravity. Once a first detection is made, a critical challenge will be to robustly extract physically interesting information from the detected signals. In this essay, we describe initial research towards answering this challenge, and thereby unleashing the full power of gravitational waves as an engine for the discovery of new physics.
\end{abstract}

\thispagestyle{empty}
\clearpage
\pagenumbering{arabic}

Neutron stars are among the densest objects in the Universe, second only to the centres of black holes. Unlike black holes, their interiors can be directly probed to study matter at extreme densities and under strong gravitational fields. As these conditions cannot be recreated in terrestrial experiments, neutron stars offer a unique opportunity to study the physics of dense nuclear matter, and to test general relativity and alternative theories of gravity in the strong-field limit.

The first theoretical prediction of neutron stars was made in 1934 by Baade and Zwicky~\cite{BaadZwic1934:RmrSprCsRy}, shortly after the discovery of the neutron. They hypothesised that neutron stars are the ultimate fate of ordinary stars once their nuclear fuel stocks are exhausted; the star collapses inward and explodes as a supernova, leaving behind a neutron star formed from its remnant stellar core. It is now known that stars with masses below about 9 solar masses will instead form white dwarf stars, whereas stars more massive than about 25 solar masses will continue to collapse into black holes~\cite{HegeEtAl2003:HMssSngStEThLf}.

Observational evidence for neutron stars arrived in 1967 with the first identification of radio pulsars by Bell and Hewish~\cite{HewiEtAl1968:ObsRpPlsRdSr}. Since their initial discovery, over 3,000 pulsars have been identified -- primarily via radio, X-ray, and gamma-ray observations -- and it is now well-established that pulsars are rapidly rotating neutron stars emitting regular pulses of electromagnetic radiation. The first discovery of a binary pulsar -- two neutron stars orbiting each other, where one star is observed as a pulsar -- was made by Hulse and Taylor in 1974~\cite{HulsTayl1975:DscPlsBnSy}. Further observations revealed that the orbit of this binary star system was decaying at a rate predicted (to near-perfect precision) by the radiation of gravitational waves due to the orbital motion of the neutron stars.

Forty-two years after this initial evidence of their existence, the first direct detection of gravitational waves was announced by the LIGO--Virgo Collaborations in 2016~\cite{Abbott2016:GW150914}. The history of gravitational waves over the century preceding this discovery was at times controversial: from Einstein's attempt in 1937 to overturn his initial 1916 prediction of their existence~\cite{Kenn2005:EnsVrPhyRv}; to pioneering, albeit premature, claims of detection by Weber in the 1960s~\cite{Webe1969:EvDscGrvRd}. Seven years on from this historic first discovery, nearly 100 gravitational wave events have been observed, all from colliding binary systems of pairs of black holes (including the first discovery), pairs of neutron stars, or a black hole and a neutron star.

Gravitational wave astronomy, despite its short history, has already made valuable contributions to our understanding of the Universe. The wealth of detected binary black hole mergers provides a unique understanding of the abundance of black holes and their observed masses, otherwise inaccessible to electromagnetic astronomy. By modelling the history of each binary system, the observed final distribution of black hole masses -- which range from a few to 150 solar masses -- is used to constrain the initial distribution of their stellar progenitors, and thereby illuminate the evolutionary processes and population dynamics of stellar binaries from birth to death.

In 2017, the spectacular first detection of gravitational waves from the merger of a binary neutron star system~\cite{Abbott2017:GW170817} was accompanied by observations across the electromagnetic spectrum by some 60 scientific collaborations~\cite{LIGOEtAl2017:MltObsBnNtSMr}. This event was the first clear association of a binary neutron star merger and a short-hard gamma-ray burst, and confirmed that neutron star mergers are responsible for gamma-ray bursts of this type. The further association of this event with a kilonova -- a bright, electromagnetic transient driven by the r-process decay of heavy elements -- demonstrates the role that neutron star collisions play in the nucleosynthesis of heavy elements in the Universe. The absolute distance to the event inferred from the gravitational wave signal alone permitted a measurement of the Hubble constant~\cite{LIGOEtAl2017:GrvStSMsrHbCn}, independent of those derived from supernovae observations and from the cosmic microwave background.

It is widely anticipated that, in the coming years, gravitational waves observations of neutron stars will provide an unrivalled laboratory for the study of extreme matter and strong gravity. Although much has been learned about neutron stars since their first discovery, much more remains to be discovered. It is presumed that neutron stars have a maximum possible mass (of likely a few solar masses) and rotational frequency (of likely a few kilohertz); the exact values are not known, however, and the physical mechanisms that control these properties are not fully understood. Ironically, despite their name implying an abundance of neutron-rich material, what exactly neutron stars are made of is also an open question. At a few kilometres below their surface, the density of neutron star matter is expected to exceed the density of normal atomic nuclei~\cite{ChamHaen2008:PhyNtrStCr}, and hence lies beyond the realm of terrestrial experiments. It has been speculated that exotic states of matter, such as hyperons, kaon condensates or deconfined quark matter, could exist at the super-nuclear densities occurring within neutron stars.

The microphysics of neutron star matter is encoded in the equation of state, which describes the relationship between physical quantities such as pressure, density and temperature. While theories abound, the neutron star equation of state is currently poorly understood. Together with an understanding of the equilibrium state of rotating stars in general relativity~\cite{Hart1967:SlRtRltSIEqtSt}, the equation of state may be inferred from the star's macroscopic properties such as its mass $M$, radius $R$, and principal moment of inertia $\Izz$ about its rotation axis. Importantly, these macroscopic properties are linked, not only to the physics of neutron star matter encoded by nuclear and quantum physics, but also to the physics of the gravitational forces of a rotating star as described by general relativity.

Observations of the maximum $M$ have been inferred from radio pulsars~\cite{DemoEtAl2010:TwsNtSMsrUsShD} and from gravitational wave observations~\cite{RezzEtAl2018:UGrvObQsnRlCnMMNS}, while measurements of $M$ and $R$ have been made using X-ray pulsars~\cite{MillEtAl2019:PJ0MRdNDImpPrpNtSM}, and of $\Izz$ from a unique double pulsar system~\cite{KramEtAl2021:StrGrvTsDbPl}. Aside from the late stages of a binary neutron star merger, where both stars are tidally deformed, these measurements are of neutron stars at an equilibrium state, where the star is expected to be close to spherically symmetric.

Gravitational wave observations may, however, be able to detect neutron stars which are deformed from spherical symmetry. A number of mechanisms, including a strong magnetic field permeating the star, accretion of matter from a companion star, or periodic oscillations akin to travelling Rossby waves in the Earth's oceans, could perturb the neutron star away from axisymmetry about its rotation axis. General relativity predicts that the neutron star would then radiate a continuous stream of gravitational waves at twice its rotational frequency; these are known as \emph{continuous gravitational waves}.

Continuous gravitational waves have yet to be detected, and there are many challenges to making a first detection. The degree of non-axisymmetric perturbation is characterised by the ellipticity $\epsilon$, which gives the fractional change in moment of inertia relative to that at sphericity; it is directly proportional to the gravitational wave amplitude. While this quantity is highly uncertain, \emph{a priori} it is expected to be very small; theoretical predictions for ``non-exotic'' matter (i.e.\ without deconfined quarks) yield $\epsilon \lesssim 10^{-6}$ at best. As a result, continuous gravitational waves are several orders of magnitude weaker than gravitational waves emitted by colliding neutron stars and black holes. While those events can be detected at billions of light years from Earth, it is expected that the terrestrial gravitational wave detectors will at best be able to detect continuous gravitational waves from neutron stars within the Milky Way galaxy (or its satellites).

Unlike the short-lived gravitational wave events detected thus far, however, continuous gravitational waves are expected to be persistent in the data, as it is assumed that the neutron star remains non-axisymmetrically perturbed over very long timescales. One can therefore apply optimal data analysis techniques which match the weak continuous gravitational wave signal against a template waveform, and thereby build up sufficient signal strength to be detectable above the background detector noise. This leads to a further challenge, however: what are the parameters of the template waveform? In some cases these parameters can be fully determined. If, for example, one suspects that a known radio pulsar is also emitting continuous gravitational waves, then one can fully determine the parameters of the continuous wave template by observing the frequency of the radio pulses over time, and deducing that gravitational waves will then be radiated at twice that frequency.

Not all neutron stars are observed as pulsars, however. Based on observed rates of supernovae, it is estimated that there may be up to a billion neutron stars in the Milky Way \cite{SartEtAl2010:GlcNtStISVlDstDH}, only a few thousand of which are currently observed as pulsars. Continuous gravitational waves offer a unique opportunity to observe those neutron stars not detectable as pulsars and consequentially invisible to electromagnetic telescopes. As the template waveform parameters of these neutron stars are unknown, one must construct a bank of templates spanning the space of possible parameters. These parameters include the position of the neutron star in the sky, and the evolution of the gravitational wave frequency over time -- one expects the neutron star to spin more slowly as its rotational energy is radiated away in gravitational waves, akin to the loss of orbital energy from the Hulse-Taylor binary pulsar. Unfortunately, this parameter space quickly becomes vast, particularly given that the continuous gravitational wave signals must be traced over a year or more of detector data. The time taken to analyse the data, even on modern computing hardware, quickly becomes prohibitive, and one must instead adopt suboptimal data analysis strategies which are less costly~\cite{TenoEtAl2021:SMtCnGrvSgUSAdvE}.

Because of the significant challenges outlined above, the question of how to make a first detection of continuous gravitational waves has received considerable attention over the last few decades. Less attention, however, has been paid to the next question: once a detection has been made, how can one gain robust insights into neutron star physics from such a detection? As the current generation of gravitational wave detectors continue to improve in sensitivity, an answer to the first question may not be far away. Should we be so fortunate, an ability to answer the second question will become vital.

Continuous gravitational waves, once detected, will provide a unique astronomical tool for examining the physics of neutron stars, complementing other techniques. While electromagnetic emission from neutron stars originates from the exterior of the star, continuous gravitational waves arise from asymmetric distributions of mass within the star, and will therefore be more sensitive to its interior physics. Furthermore, while electromagnetic observations infer macroscopic properties of spherically symmetric neutron stars, and binary neutron star merger signals last but a short time, continuous gravitational waves will probe the non-symmetric state of deformed neutron stars over long periods. It is therefore critical that, in anticipation of a first detection, data analysis methods are developed that can extract the physical information conveyed by continuous gravitational waves in a robust manner.

Initial research into this question by Sieniawska and Jones~\cite{SienJone2022:GrvWSpNtSNtqSr} demonstrated that, in fact, extracting this information is far from straightforward. Two equations govern the properties of a continuous gravitational wave signal: an equation for its characteristic amplitude $h_0$, derived ultimately from the famous ``quadrupole formula'' for gravitational waves; and an equation for its frequency $f$, derived from a conservation of energy argument that the energy radiated in gravitational waves is fed by the loss of rotational energy of the star. These equations, however, contain three unknowns: the principal moment of inertia $I_{zz}$, ellipticity $\epsilon$, and distance to the star $\dist$. This illustrates that, unlike for binary neutron star/black hole collisions, the distance to a source of continuous gravitational waves cannot be determined absolutely, and must instead be determined independently in order to break the degeneracy. One proposal is to observe the parallax of sufficiently close-by neutron stars and its effect on the continuous gravitational wave signal to determine the distance~\cite{SienEtAl2023:MsNtrDsPrGrvP}.

\begin{figure}[!t]
    \centering
    \includegraphics[width=0.9\linewidth]{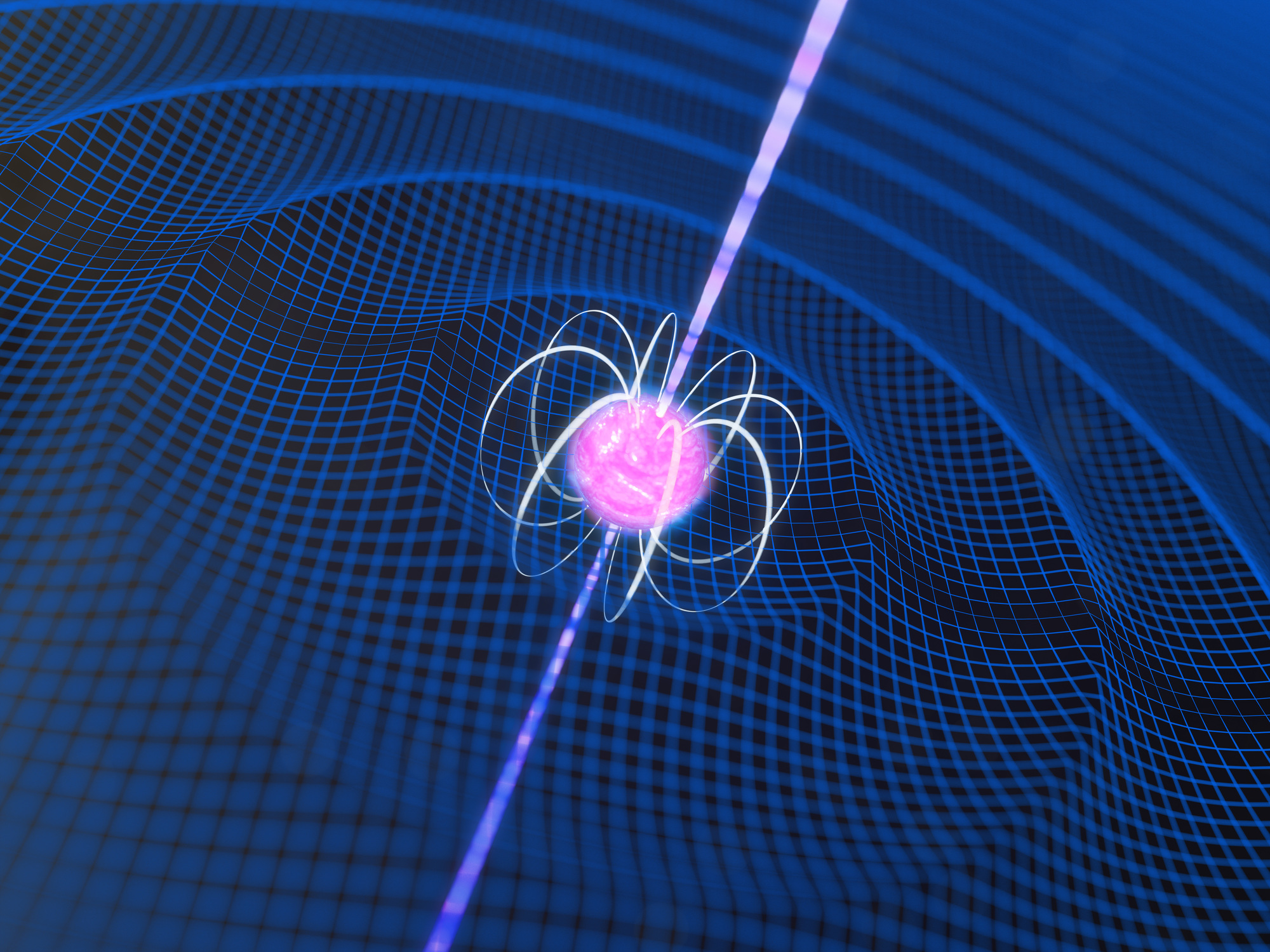}
    \caption{Artist's impression of a neutron star radiating both continuous gravitational waves and electromagnetic radiation. Credit: OzGrav/Carl Knox.}
    \label{fig:NS_with_CW_EM}
\end{figure}

In a recent paper~\cite{Lu2023}, we address this degeneracy by instead assuming that the neutron star radiates both continuous gravitational waves and electromagnetic radiation (Figure~\ref{fig:NS_with_CW_EM}). Aside from providing important corroboration to a continuous gravitational wave detection, an electromagnetic counterpart would likely provide an independent measurement of the neutron star distance. Two possibilities are: the detection of a radio pulsar, which would provide a distance measurement from the dispersion of the pulses as they travel through the interstellar medium; or the detection of a supernova remnant, where a distance could be determined from observations of the radial expansion of the remnant.

The assumption that the rotational energy of the neutron star is lost through electromagnetic waves, in addition to continuous gravitational wave radiation, introduces another unknown parameter: the strength of the magnetic field of the star, which drives its electromagnetic radiation. It also introduces an additional constraint, however: the braking index $n$. This quantity is a measure of which mode of radiation dominates the energy lost by the star: pure electromagnetic radiation implies $n = 3$, while pure gravitational wave radiation implies $n = 5$. The braking index may be deduced by measuring the second time derivative $\fddot$ of the gravitational wave frequency.

The additional constraint provided by the braking index $n$, and the assumption of a measured distance $\dist$ to the star, permits the following three macroscopic properties of a neutron star to be completely determined: $I_{zz}$, representing the spherical equilibrium state of the star; $\epsilon$, representing its perturbation away from axisymmetry; and $\pmp$, the component of the star's magnetic moment perpendicular to its rotation axis. A continuous gravitational wave detection determines its characteristic amplitude $h_0$, frequency $f$, and its time derivatives $\fdot$ and $\fddot$, which then determine $n$. The three properties are then given by the following relations (omitting known physical constants for brevity):
\begin{align*}
  \Izz &\propto \frac{ \dist^2 h_0^2 f }{ \fdot ( 3 - n )} \, &
  \epsilon &\propto \frac{ \fdot ( 3 - n ) }{ \dist h_0 f^3 } \, &
  \pmp &\propto \frac{ \dist h_0 }{ f } \sqrt{ \frac{ n - 5 }{ 3 - n } } \,.
\end{align*}
These simple relations already yield some key insights: $\Izz$ and $\epsilon$ cannot be determined in the absence of continuous gravitational waves (which implies $n = 3$), whereas $\pmp$ requires emission of both electromagnetic and gravitational wave radiation (which implies $3 < n < 5$) to be determinable.

\begin{figure}[!t]
    \centering
    \includegraphics[width=0.7\linewidth]{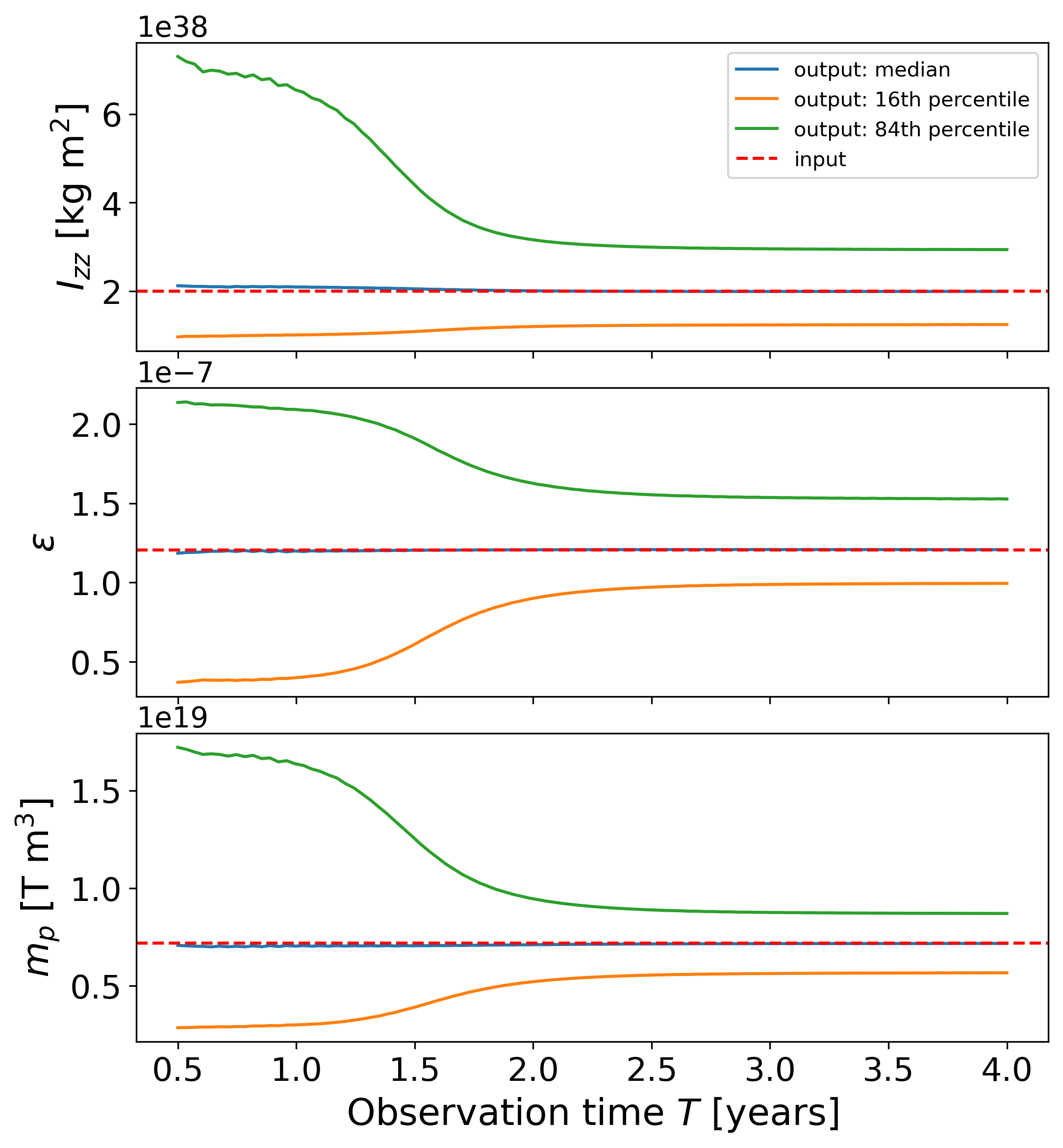}
    \caption{Simulated example of the measured (output) values of $\Izz$, $\epsilon$, $\pmp$, and their errors, compared to their true (input) values, after $T$ years of gravitational wave observations. Reproduced from Figure~1 of~\cite{Lu2023}.}
    \label{fig:Single_NS}
\end{figure}

Of course, gravitational wave detectors are not perfect apparatuses. Noise arising from a wide variety of sources -- instrumental, environmental, and anthropogenic -- will impair the precise measurement of the continuous gravitational wave parameters, and hence the inference of $\Izz$, $\epsilon$, and $\pmp$. It is vital, therefore, to robustly characterise to what accuracy we might expect to measure these parameters. In~\cite{Lu2023} we perform a first attempt at characterising the expected errors, based on a simple approach using the Fisher information matrix.

An example result simulating the expected errors in the inferred $\Izz$, $\epsilon$, and $\pmp$ is reproduced in Figure~\ref{fig:Single_NS}. It suggests that, after only a few years of observing continuous gravitational waves, the errors will converge to a minimum value (determined by the error in measuring the stellar distance $\dist$). Further simulations suggest that, modulo a number of caveats detailed in~\cite{Lu2023}, we could achieve errors in $\Izz$ of $\sim 32\%$ and errors in $\epsilon$ and $m_p$ of $\sim 16 \%$.

This degree of accuracy demonstrates that continuous gravitational waves have much to offer the study of neutron stars. The error in $\Izz$ is competitive with other techniques~\cite{MiaoEtAl2022:MmInPJ07LIGNI}, and has the important advantage of not relying on an \emph{a priori} assumption of the neutron star equation of state. A measurement of $\pmp$ would provide an independent confirmation of a common formula used to infer the magnetic field strength of pulsars~\cite[e.g.\ ][]{CondRans2016:EssRdAst}. A measurement of $\epsilon$, however, is unique to continuous gravitational waves, and is therefore an unprecedented probe into the state of matter in an asymmetrically perturbed neutron star.

Our sensitivity to gravitational waves continues to improve. Upgrades to the existing, second-generation gravitational wave detectors are ongoing, and it is anticipated that these detectors will eventually begin operating near-continuously, vastly increasing the amount of data available for analysis. Planning is also well underway for a third generation of ground-based gravitational wave detectors, to begin construction in the 2030s, with orders of magnitude boosts to sensitivity expected.

The promise of gravitational wave detection has always been to open a new window onto the Universe. A century on from their prediction by Einstein, and a mere seven years from their first discovery, gravitational waves are firmly established as an essential tool of modern astrophysics. New astronomical instruments have long led to new discoveries -- from Galileo's first observations of the moons of Jupiter, to the inadvertent discovery of the cosmic microwave background -- and we may be optimistic that gravitational waves will follow in that mould.

A first detection of continuous gravitational waves would be an exceptional scientific discovery. But what we learn from that discovery could be even more extraordinary. Together with observations of pulsars and colliding neutron stars, we may one day use continuous gravitational waves to piece together answers to the following fundamental questions:

What are neutron stars made of? Are they really made only of neutrons and ordinary particles? Or do more exotic states of matter lie within?

\end{document}